\documentclass[12pt]{iopart}

\usepackage{amsfonts}
\usepackage{mathrsfs}
\usepackage{color}
\usepackage{hyperref}
\usepackage{cite}
\usepackage{float}
\usepackage{graphicx}
\usepackage{graphics}
\def\R{{\mathbb{R}}}

\def\C{\mathbb{C}}
\def\Z{\mathbb{Z}}
\def\Q{\mathbb{Q}}

\def\H{{\mathcal{H}}}
\date{\empty}
\def\<{\langle}
\def\>{\rangle}
\def\tr{{\rm{tr}}}
\def\i{{\rm{i}}}
\def\e{\rm{e}}

\newcommand{\be}{\begin{equation}}
\newcommand{\ee}{\end{equation}}
\newcommand{\bea}{\begin{eqnarray}}
\newcommand{\eea}{\end{eqnarray}}
\newcommand{\bna}{\begin{eqnarray*}}
\newcommand{\ena}{\end{eqnarray*}}
\renewcommand{\theequation}{\thesection.\arabic{equation}}

\begin{document}
\title[Masking quantum information]{Masking quantum information encoded in pure and mixed states}

\author{Huaixin Cao, Yuxing Du, Zhihua Guo, Kanyuan Han and Chuan Yang}

\address{School of Mathematics and Information Science,  Shaanxi Normal University, Xi'an 710119, China}
\ead{guozhihua@snnu.edu.cn, caohx@snnu.edu.cn}
\vspace{10pt}

\begin{abstract}
Masking of quantum information means that information is hidden from a subsystem and spread over a composite system. Modi et al. proved in [Phys. Rev. Lett. 120, 230501 (2018)] that this is true for some restricted sets of nonorthogonal quantum states and it is not possible for arbitrary quantum states.
In this paper, we discuss the problem of masking quantum information encoded in pure and mixed states, respectively. Based on an  established necessary and sufficient condition for
a set of pure states to be masked by an operator, we find that there exists a set of four states that can not be masked, which implies that to mask unknown pure states is impossible. We  construct a masker $S^\sharp$ and obtain its maximal maskable set, leading to an affirmative answer to a conjecture proposed in Modi's paper mentioned above. We also prove that an orthogonal (resp. linearly independent) subset of pure states can be masked by an isometry (resp. injection). Generalizing the case of pure states, we introduce the maskability of a set of mixed states and prove that a commuting subset of mixed states can be masked by an isometry $S^{\diamond}$ while it is impossible to mask all of mixed states  by any operator. We also find the maximal maskable sets of mixed states of the isometries ${S^{\sharp}}$ and ${S^{\diamond}}$, respectively.
\end{abstract}




\section{Introduction}
\setcounter{section}{1}

In quantum mechanics, there are many ``no-go theorems" meaning that to do something according to quantum theory is impossible,
say the no-cloning theorem \cite{Clon[1],Clon[2],Clon[3],Clon[4],Clon[5]}, the no-broadcasting
theorem \cite{broad[5],broad[6]}, the no-deleting theorem \cite{delt1,delt2}, the no-hiding Theorem\cite{hiding[7]}, and
no-signalling theorem \cite{sign}.

To deal with the encoding of quantum
information in an arbitrary composite quantum state, Modi et al. \cite{Mask} discussed the problem of masking quantum information contained in some pure states with a linear operator and obtained the so called no-masking theorem, which says that it is impossible to mask arbitrary states by the same operator. It was also proved in  \cite{Mask} that there are sets of nonorthogonal states whose information can be masked. Just as no-go theories being of great significance in information
processing \cite{[13],[14],[18],18}, masking of quantum information has potential applications\cite{[22],[24]}. Li and Wang \cite{Li-Wang} discussed the problem of masking quantum information in multipartite scenario and proved that quantum states can be masked when more
participants are allowed in the masking process. Li et al. \cite{LiBo1}
considered the problem of what kinds of quantum states can be either deterministically or probabilistically masked and proved that mutually orthogonal quantum states can always be served for deterministic masking of quantum information. They also  constructed a probabilistic masking machine for linearly independent states. Liang et al.  \cite{LiBo2} studied the problem of information masking through nonzero linear operators and proved that a nonzero linear operator cannot mask any nonzero measure set of qubit states. They also shown that the maximal maskable set of states on the Bloch
sphere with respect to any masker is the ones on a spherical circle. Furthermore, they given a proof of the conjecture on maskable qubit states proposed by Modi et al. in \cite{Mask}. Moreover, Li and Modi \cite{LiMoa} discussed the problems of probabilistic and approximate masking of quantum
information and the performance of a masking protocol when we are allowed (probabilistic) approximate protocol. They also proved that an $\varepsilon$-approximate universal masker for all states does not exist if the error bound $\varepsilon$ is less than a bound.
Ding and Hu discussed in \cite{Ding} masking quantum information on hyperdisks and the structure of the set of maskable states, Lei et. al.  studied unconditionally secure qubit commitment scheme by using quantum maskers \cite{LieK} and randomness cost of masking
quantum information and the information conservation law \cite{LieJ}, Ghoshet. al. \cite{Ghosh} pointed out that it is possible for classical information encoded in composite quantum states to be completely masked from reduced sub-systems.

In this paper, we continue to discuss the problem of masking quantum information encoded in pure and mixed states, respectively, including the mathematical definitions, characterizations, masking theorems and no-masking theorems. In Section 2, we will recall and redefine of the maskability of a set of pure states, derive a
necessary and sufficient condition for
a set of pure states to be masked by an operator, find a set of four states that can not be masked, which implies that to mask unknown pure states is impossible. We construct a masker $S^\sharp$ and obtain its maximal maskable set, and then obtain an affirmative answer to Conjecture 5 in \cite{Mask}. We also prove by definition that an orthogonal subset of pure states can be masked by an isometry. In Section 3, by generalizing the case of pure states, we will introduce the maskability of a set of mixed states and prove that a commuting subset of mixed states can be masked by an isometry $S^{\diamond}$ while it is impossible to mask all of mixed states. We also find the maximal maskable sets of the maskers  $S^{\sharp}$ and $S^{\diamond}$, respectively.

\section{Masking of pure states}
\setcounter{section}{2}
We use notations $PS_X$ and $D_X$ to denote the sets of all pure states (unit vectors in the Hilbert space $\H_X$) and all mixed states (density operators on  $\H_X$) of a quantum system $X$, and use $B(\H_X)$ to denote the set of all bounded linear operators on $\H_X$. We also use $[n]$ to denote the set $\{1,2,\ldots,n\}.$

Let us rewrite the definition of a masker introduced by Modi et al. in \cite{Mask}.

{\bf Definition 2.1.} Let $Q$ be a subset of $PS_A$ and  $S:\H_A\rightarrow \H_A\otimes\H_B$ be a linear operator. If there are mixed states $\rho_A\in D_A,\rho_B\in D_B$ such that
\be\label{D1}
\tr_B[S|\psi\>\<\psi|S^\dag]=\rho_A{\rm{\ and\ }}
\tr_A[S|\psi\>\<\psi|S^\dag]=\rho_B,\ \  \forall |\psi\>\in Q,\ee
then we say that the information contained in $Q$ is masked by $S$; shortly, $Q$ is masked by $S$, or $Q$ is a maskable set of $S$.
We also say that the operator $S$ is a {\it quantum
information masker} (shortly, a masker) for $Q$ and that $\H_B$ is a {\it masking space} for $Q$.

When the space $\H_B$ and the operator $S$ satisfying the masking  conditions (\ref{D1}) exist, we say that $Q$ can be masked.

See Fig. 1 for the definition above.

\begin{figure}[H]
\centering
\includegraphics[width=12cm,height=4cm]{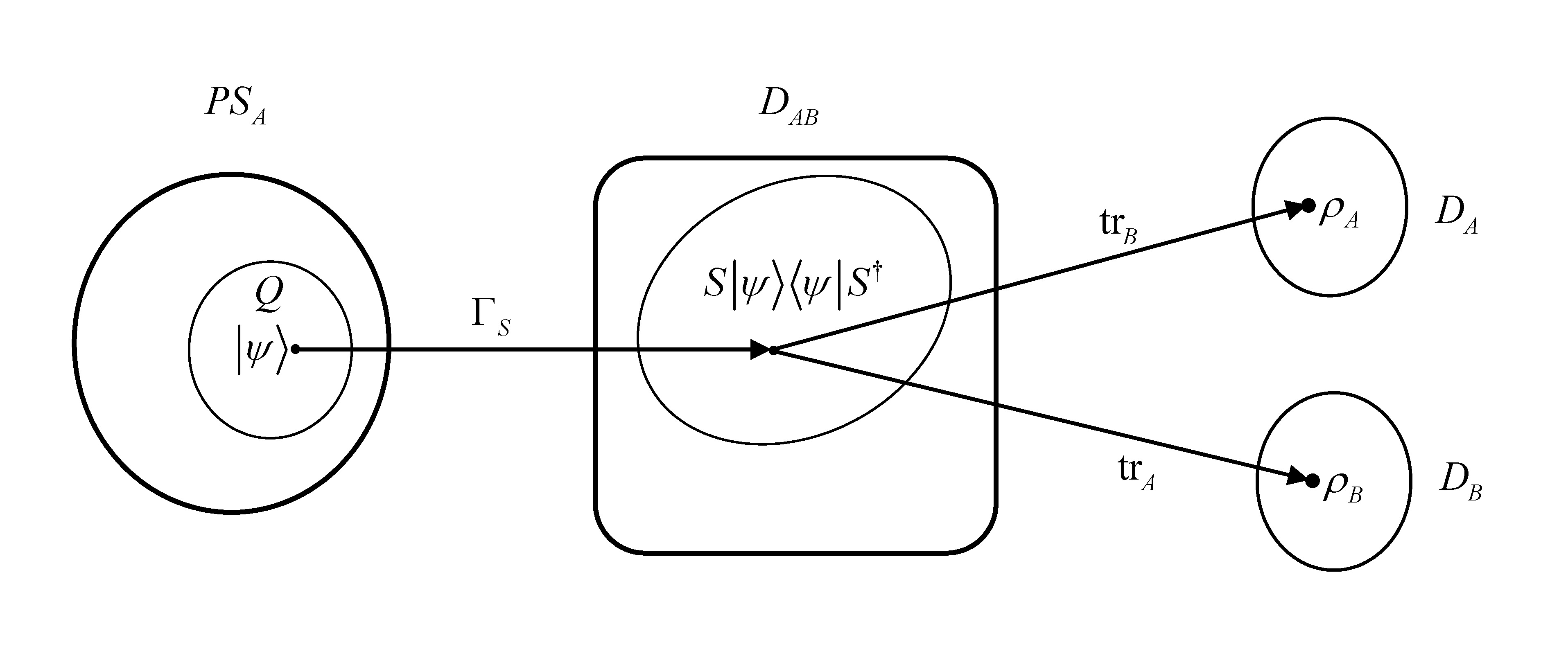}%
\caption{An illustration of Definition 2.1, in which  $\Gamma_S(|\psi\>)=
S|\psi\>\<\psi|S^\dag$ and all pure states in $Q$ are mapped to the fixed states $\rho_A$ and $\rho_B$ under the composite maps $\tr_B\circ\Gamma_S$ and $\tr_A\circ\Gamma_S$, respectively. Thus, the quantum information contained in the states from $Q$ is masked by $S$.}
\label{Figure1}
\end{figure}

{\bf Remark 2.1.} Here are some observations for a masker, which were partially mentioned in \cite{Mask}.

(1) By Definition 2.1, a masker $S:\H_A\rightarrow \H_A\otimes\H_B$ for a set $Q$ transforms pure states $|\psi\>$ in $Q$ as  pure states  $S|\psi\>$ such that density matrices $S|\psi\>\<\psi|S^\dag$ have the same marginal states $\rho_A$ and $\rho_B$. Thus, a natural condition for a masker $S$ should be $S^\dag S=I_A$, i.e.  $S$  is an isometry. We call such an $S$ an {\it isometric masker}.

(2) To model physically an isometric masker $S:\H_A\rightarrow \H_A\otimes\H_B$ for a set $Q\subset PS_A$ with a unitary operator $U_S$ on the Hilbert space $\H_A\otimes\H_B$, we first fix a pure state $|b\>$ in the ancillary system $B$ and define a unitary operator (a surjective isometry) $\tilde{S}:\H_A\otimes\C|b\>\rightarrow {\rm{ran}}(S)$ as $\tilde{S}(|\psi\>|b\>)=S|\psi\>$. Then we choose a unitary operator $V:\H_A\otimes|b\>^\perp\rightarrow \ker(S^\dag)$ and define an operator $U_{S,V}:\H_A\otimes\H_B\rightarrow \H_A\otimes\H_B$ by
$$U_{S,V}=\left(
        \begin{array}{cc}
          \tilde{S} & 0 \\
          0 & V \\
        \end{array}
      \right): (\H_A\otimes\C|b\>)\oplus (\H_A\otimes|b\>^\perp)\rightarrow {\rm{ran}}(S)\oplus \ker(S^\dag).$$
Clearly, $U_{S,V}$ is a unitary operator satisfying
$$U_{S,V}(|\psi\>|b\>)=\tilde{S}(|\psi\>|b\>)=S|\psi\>,\ \ \forall |\psi\>\in Q.$$
This shows that an isometric masker $S:\H_A\rightarrow \H_A\otimes\H_B$ for a set $Q\subset PS_A$ can be modeled by a unitary operator $U_{S,V}:\H_A\otimes\H_B\rightarrow \H_A\otimes\H_B$ in such a way that
$$S|\psi\>=U_{S,V}(|\psi\>|b\>),\ \ \forall |\psi\>\in Q.$$

(3) When the information contained in $Q$ is masked by $S$, the information contained in  $Q\cup\{|\psi\>\}$  is also masked by $S$ provided that the density matrix $|\psi\>\<\psi|=\sum_{k=1}^mc_k|\psi_k\>\<\psi_k|$ for some $|\psi_1\>,|\psi_2\>,\ldots, |\psi_m\>$ in $Q$. Hence, the information contained in
$$\tilde{Q}:=Q\cup\left\{|\psi\>\in PS_A: |\psi\>\<\psi|=\sum_{k=1}^mc_k|\psi_k\>\<\psi_k|, c_k\in\C,|\psi_k\>\in Q(\forall k\in[m])\right\}$$
 is also masked by $S$.

(4) When  $Q$ is masked by an isometry $S$, there exists at least one state $|\psi\>$ in $Q$ such that $S|\psi\>$ is entangled unless  $Q$ has just one state. Indeed, if for every $|\psi\>$ in $Q$, $S|\psi\>$ is separable, say $S|\psi\>=|f_\psi\>|g_\psi\>$, then $|f_\psi\>\<f_\psi|=\rho_A$ and $|g_\psi\>\<g_\psi|=\rho_B$  for every $|\psi\>$ in $Q$. Taking a fixed $|\psi_0\>\in Q$ yields that
$|f_\psi\>\<f_\psi|=|f_{\psi_0}\>\<f_{\psi_0}|$ and $|g_\psi\>\<g_\psi|=|g_{\psi_0}\>\<g_{\psi_0}|$ and so $|f_\psi\>=e^{i\alpha_{\psi}}|f_{\psi_0}\>$ and $|g_\psi\>=e^{i\beta_{\psi}}|g_{\psi_0}\>$ for all  $|\psi\>$ in $Q$. Thus,
$$S|\psi\>=e^{i(\alpha_{\psi}+\beta_{\psi})}|f_{\psi_0}\>|g_{\psi_0}\>=e^{i(\alpha_{\psi}+\beta_{\psi})}S|\psi_0\>
=S\left(e^{i(\alpha_{\psi}+\beta_{\psi})}|\psi_0\>\right)$$  for all  $|\psi\>$ in $Q$. Since $S$ is injective, we see that $|\psi\>=e^{i(\alpha_{\psi}+\beta_{\psi})}|\psi_0\>$  for all  $|\psi\>$ in $Q$. Physically,  $Q$ contains just one state $|\psi_0\>$. From the discussion here, we also see that  when  $Q$ is masked by an isometry $S$, all image states $S|\psi\>$ are entangled unless  the marginal states both $\rho_A$ and $\rho_B$ are pure states.

The maskability of an orthogonal set of pure states was proved in  \cite[Theorem 1]{LiBo1} in light of their Lemma 1. However, that lemma is indeed a necessary (not sufficient) condition for a set to be masked (Remark 2.4, below).  Next, we give a direct proof of this result.

{\bf Proposition 2.1.} {\it Any orthogonal set $Q$ in $PS_A$ can be masked by an isometry from $\H_A$ into $\H_A\otimes\H_B$ with $\H_B=\H_A$.}

{\bf Proof.} Let $Q=\{|\psi_1\>,|\psi_2\>,\ldots,|\psi_n\>\}$ be an orthogonal set in  $PS_A$. Clearly, we may assume that $Q$ is an orthonormal basis for $\H_A$. Denoted by $F_n=[c_{kj}]$ the quantum Fourier transform of order $n$, i.e. $c_{kj}=\frac{1}{\sqrt{n}}\omega_n^{(k-1)(j-1)}$ where $\omega_n=e^{2\pi\i/n}$, which is an $n$ by $n$ unitary matrix, and define
\be\label{OM}
S_{F_n}|\psi_i\>=\sum_{j=1}^nc_{ij}|\psi_j\>|\psi_j\>(i=1,2,\ldots,n).
\ee
Since
$$\<\psi_i|S_{F_n}^\dag S_{F_n}|\psi_k\>=\sum_{j=1}^n\overline{c_{ij}}{c_{kj}}=\delta_{j,k},$$
$S_{F_n}$ can be linearly extended as an isometry from $\H_A$ into $\H_A\otimes\H_B$ where $\H_B=\H_A$. Put
$$\rho_A=\rho_B=\frac{1}{{n}}\sum_{j=1}^n|\psi_j\>\<\psi_j|,$$
then $\rho_A\in D_A, \rho_B\in D_B.$
For all $i=1,2,\ldots,n$, we compute that
\bna
\tr_A(S_{F_n}|\psi_i\>\<\psi_i|S_{F_n}^\dag)&=&
\sum_{j,k=1}^nc_{ij}\overline{c_{ik}}\tr_A\left(|\psi_j\>\<\psi_k|\otimes
|\psi_j\>\<\psi_k|\right)\\
&=&\sum_{j=1}^n|c_{ij}|^2|\psi_j\>\<\psi_j|\\
&=&\rho_B.
\ena
Similarly,
$$\tr_B(S_{F_n}|\psi_i\>\<\psi_i|S_{F_n}^\dag)=\rho_A$$
for all $i=1,2,\ldots,n$. By Definition 2.1, $Q$ can be masked by $S_{F_n}$. The proof is completed.

Proposition 2.1 means that an orthogonal subset $Q=\{|\psi_1\>,|\psi_2\>,\ldots,|\psi_n\>\}$ of $PS_A$ can be masked by the operator $S_{F_n}$ defined by (\ref{OM})with the marginal states $\rho_A$ and $\rho_B$. An interesting question is whether $Q$ is the maximal maskable set of
 $S_{F_n}$. To discuss the answer to this question, we let $|\psi\>=a_1|\psi_1\>+a_2|\psi_2\>$ with $a_1>0,a_2\ne0, |a_1|^2+|a_2|^2=1$. Then
$$\tr_B(S_{F_n}|\psi\>\<\psi|S^\dag)=\rho_A+2a_1{\rm{Re}}\left(
\overline{a_2}\sum_{j=1}^nc_{1j}\overline{c_{2j}}|\psi_j\>\<\psi_j|\right).$$
Since
$c_{1j}\overline{c_{2j}}=\frac{1}{n}\omega_n^{-j+1}$, we get
$$T:=\overline{a_2}\sum_{j=1}^nc_{1j}\overline{c_{2j}}|\psi_j\>\<\psi_j|=\frac{1}{n}\overline{a_2}
\sum_{j=1}^n\omega_n^{-j+1}|\psi_j\>\<\psi_j|,$$
and so
$$2{\rm{Re}}(T)={T+T^\dag}=\frac{1}{n}\sum_{j=1}^n\left(\overline{a_2}
\omega_n^{-j+1}+a_2\omega_n^{j-1}\right) |\psi_j\>\<\psi_j|.$$
Thus, $\tr_B(S_{F_n}|\psi\>\<\psi|S_{F_n}^\dag)=\rho_A$ if and only if ${\rm{Re}}(a_2\omega_n^{j-1})=0$ for all $j=1,2,\ldots,n.$ In the case that $n=2$, $\tr_B(S_{F_2}|\psi\>\<\psi|S_{F_2}^\dag)=\rho_A$ holds if and only if $a_2$ is an imaginary number; in the case that $n>2$, $\tr_B(S_{F_n}|\psi\>\<\psi|S_{F_n}^\dag)\ne\rho_A$.
As a conclusion, when $n=2$, the operator $S_{F_2}$ defined by (\ref{OM}) can mask not only the orthogonal set $Q=\{|\psi_1\>,|\psi_2\>\}$ but also the nonorthogonal set
$$Q_1=\{|\psi_1\>,|\psi_2\>\}\cup\{a|\psi_1\>+ib|\psi_2\>: a,b\in\R, a^2+b^2=1\},$$
but it can not mask the set
$$Q_2=\{|\psi_1\>,|\psi_2\>\}\cup\{a_1|\psi_1\>+a_2|\psi_2\>: a_1>0,a_2\ne\R\i, |a_1|^2+|a_2|^2=1\}.$$
When $n>2$, the operator $S_{F_n}$ defined by (\ref{OM}) can mask the orthogonal set $Q=\{|\psi_1\>,|\psi_2\>,\ldots, |\psi_n\>\}$, but it can not mask any nonorthogonal set
$$Q_1=\{|\psi_1\>,|\psi_2\>,\ldots,|\psi_n\>,a_1|\psi_1\>+a_2|\psi_2\>\},$$
where $a_1,a_2\in\C\setminus\{0\},|a_1|^2+|a_2|^2=1$.

 Li and Wang \cite[Theorem 1]{Li-Wang} proved that when $n\ge2$,  all the quantum states in $\C^n$  can be masked into $\bigotimes_{j=1}^{2n}\H_{A_j}$ where $\H_{A_j}=\C^n.$ Here is an alternative proof of this result in light of the operator $S_{F_n}$ defined  by (\ref{OM}). Choose an orthonormal basis $\{|i\>\}_{i=1}^n$ for $\C^n$, put $c_{kj}=\frac{1}{\sqrt{n}}\omega_n^{(k-1)(j-1)}$ and put
$$|\psi_k\>=S_{F_n}|k\>=\sum_{j=1}^nc_{kj}|j\>|j\>(k=1,2,\ldots,n),$$
$$S|k\>=|\Psi_k\>:=\bigotimes_{j=1}^{n}|\psi_k\>=\left(\frac{1}{\sqrt{n}}\right)^n\sum_{j_1,j_2,\ldots,j_n\in[n]}\omega_n^{k|\textbf{j}|}|j_1j_1j_2j_2\cdots j_nj_n\>$$
for all $k=1,2,\ldots,n$ where $|\textbf{j}|=j_1+j_2+\cdots+j_n$. Since
$$\<\Psi_k|\Psi_j\>=\prod_{j=1}^{n}\<\psi_k|\psi_j\>=\delta_{k,j},$$
 $S$ can be linearly extended as an isometry from $\H_{A_1}$ into $\bigotimes_{j=1}^{2n}\H_{A_j}$.
For every state $|\psi\>=\sum_{k=1}^na_k|k\>$ in $\H_{A_1}$, we compute that
$$S|\psi\>=\left(\frac{1}{\sqrt{n}}\right)^n\sum_{j_1,j_2,\ldots,j_n\in[n]}
\sum_{k=1}^na_k\omega_n^{k|\textbf{j}|}|j_1j_1j_2j_2\cdots j_nj_n\>$$
and so
$$
S|\psi\>\<\psi|S^\dag=\left(\frac{1}{n}\right)^n\sum_{j_x,i_x\in[n](x\in[n])}
\sum_{k=1}^n\sum_{\ell=1}^na_k\overline{a_\ell}\omega_n^{k|\textbf{j}|-\ell|\textbf{i}
|}\cdot|j_1j_1j_2j_2\cdots j_nj_n\>\<i_1i_1i_2i_2\cdots i_ni_n|.
$$
Put $\hat{A_j}=\{A_1,A_2,\ldots,A_{2n}\}\setminus\{A_j\}$, then
$$
\tr_{\hat{A_1}}[S|\psi\>\<\psi|S^\dag]=\left(\frac{1}{n}\right)^n\sum_{j_1\in[n]}
\sum_{k=1}^n\sum_{\ell=1}^na_k\overline{a_\ell}\omega_n^{kj_1-\ell j_1}\sum_{j_2,\ldots,j_n\in[n]}\omega_n^{(k-\ell)\sum_{t=2}^nj_t}|j_1\>\<j_1|.$$
Note that
$$\sum_{m=1}^n\omega_n^{pm}=\frac{1-(\omega_n^n)^p}{1-\omega_n^{p}}=0$$
for every nonzero integer $p$, we see that
$$\sum_{j_2,\ldots,j_n\in[n]}\omega_n^{(k-\ell)\sum_{t=2}^nj_t}=\prod_{t=2}^n\sum_{j_t=1}^n\omega_n^{(k-\ell)j_t}=0(k\ne\ell),$$
it is $n^{n-1}$ if $k=\ell$. Thus,
$$\rho_{A_1}=\tr_{\hat{A_1}}[S|\psi\>\<\psi|S^\dag]=\frac{1}{n}\sum_{j_1\in[n]}
\sum_{k=1}^n|a_k|^2|j_1\>\<j_1|=\frac{1}{n}I_n,$$
since $\sum_{k=1}^n|a_k|^2=1$. By the symmetry of the $2n$ systems, we have  $\rho_{A_j}=\rho_{A_1}=\frac{1}{n}I_n$ for all $j=1,2,3,\ldots,2n.$ It follows from \cite[Definition 1]{Li-Wang} that all the quantum states in $\C^n$  can be masked into $\bigotimes_{j=1}^{2n}\H_{A_j}$ by the operator $S$.

The following example says that an operator may mask an infinite number of sates.

{\bf Example 2.1.} Let  $\{|k\>\}_{k=1}^d$ be an orthonormal basis (ONB) for $\H_A$ and $S$ a linear isometry from $\H_A$ into $\H_A\otimes\H_A$ with
$S:|k\>\mapsto |k\>|k\>(\forall k=1,2,\ldots,d)$. Put
$$Q=\left\{\frac{1}{\sqrt{d}}\sum_{k=1}^d{\e}^{{\i}\phi_k}|k\>:\phi_k\in(-\pi,\pi](\forall k\in[d])\right\}.$$

For any $|\psi\>=\frac{1}{\sqrt{d}}\sum_{k=1}^d{\e}^{\i x_k}|k\>$ in $Q$, we compute that
$$S|\psi\>\<\psi|S^\dag=\frac{1}{d}\sum_{k=1}^d
\sum_{j=1}^d{\e}^{{\i}(x_k-x_j)}|k\>\<j|\otimes|k\>\<j|$$
and so
$$\tr_B(S|\psi\>\<\psi|S^\dag)=\frac{1}{d}\sum_{k=1}^d|k\>\<k|=\frac{1}{d}I_A\in D_A.$$ Similarly,
$$\tr_A(S|\psi\>\<\psi|S^\dag)=\frac{1}{d}\sum_{j=1}^d|j\>\<j|=\frac{1}{d}I_B\in D_B.$$
It follows from Definition 2.1 that  $Q$ can be masked by $S$. Clearly, $Q$ is not an orthogonal set.

In order to prove a characterization of a masker (Theorem 2.1), we need the following two lemmas on the constructions of two purifications of a mixed state, which may be well-known. But, for the readers' convenience, we will give their proofs.

{\bf Lemma 2.1.} {\it Let two pure states $|\psi^{AB}_1\>$ and $|\psi^{AB}_2\>$ be purifications of a mixed state $\rho_A$ of system $A$. Then  $|\psi^{AB}_1\>$ and $|\psi^{AB}_2\>$ have the following Schmidt decompositions:
\be\label{24}
|\psi^{AB}_1\>=\sum_{j=1}^r{\sqrt{c_j}}|e_j\>|f_j\>, |\psi^{AB}_2\>=\sum_{j=1}^r{\sqrt{c_j}}|e_j\>|g_j\>
\ee
where $c_j(j=1,2,\ldots,r)$ are positive numbers, independent of $|\psi^{AB}_1\>$, the sets $\{|e_j\>\}_{j=1}^r$ and $\{|g_j\>\}_{j=1}^r$ are orthonormal  in $\H_A$ and $\H_B$, respectively,  independent of $|\psi^{AB}_1\>$, and the set $\{|f_j\>\}_{j=1}^r$ is orthonormal in $\H_B$, depending on both $|\psi^{AB}_1\>$ and $|\psi^{AB}_2\>$.}

{\bf Proof.} Let
\be\label{25}
|\psi^{AB}_1\>=\sum_{k=1}^{r'}{\sqrt{c'_k}}|e'_k\>|f'_k\>{\rm{\ and \ }} |\psi^{AB}_2\>=\sum_{j=1}^{r''}{\sqrt{c''_j}}|e''_j\>|f''_j\>
\ee
be Schmidt decompositions of $|\psi^{AB}_1\>$ and  $|\psi^{AB}_2\>$, respectively, where $c'_k,c''_j>0(\forall k,j)$, $\{|e'_k\>\}_{k=1}^{r'}$ and $\{|e''_j\>\}_{j=1}^{r''}$ are orthonormal sets in $\H_A$, $\{|f'_k\>\}_{k=1}^{r'}$ and $\{|f''_j\>\}_{j=1}^{r''}$ are orthonormal sets.
Since both $|\psi^{AB}_1\>$ and $|\psi^{AB}_2\>$ are two purifications of  $\rho_A$, we have $\tr_B|\psi^{AB}_1\>\<\psi^{AB}_1|=\tr_B|\psi^{AB}_2\>\<\psi^{AB}_2|=\rho_A$, that is,
\be\label{(26)}
\sum_{k=1}^{r'}{{c'_k}}|e'_k\>\<e'_k|= \sum_{j=1}^{r''}{c''_j}|e''_j\>\<e''_j|=\rho_A.
\ee
Thus, $r'=r''=\dim({\rm{ran}}(\rho_A))$, we denote it by $r$. Choose an $r$ by $r$ unitary matrix $U=[u_{ij}]$ such that
$|e'_k\>=\sum_{j=1}^ru_{kj}|e''_j\>(k=1,2,\ldots,r)$. Hence,
\be\label{(27)}
|\psi^{AB}_1\>=\sum_{k=1}^{r}\sum_{j=1}^r {\sqrt{c'_k}}u_{kj}|e''_j\>|f'_k\>=\sum_{j=1}^r{\sqrt{c''_j}}|e''_j\>|f_j\>
\ee
where $$|f_j\>=\frac{1}{{\sqrt{c''_j}}}\sum_{k=1}^{r} {\sqrt{c'_k}}u_{kj}|f'_k\>(j=1,2,\ldots,r).$$
It follows from Eqs. (\ref{(26)}) and  (\ref{(27)}) that
$$\sum_{j=1}^{r}{c''_j}|e''_j\>\<e''_j|=\rho_A=
\sum_{j=1}^{r}\sum_{k=1}^{r}{\sqrt{c''_jc''_k}}\<f_j|f_k\>\cdot  |e''_k\>\<e''_j|.$$
This implies that $\<f_j|f_k\>=\delta_{k,j}$ for all $k,j=1,2,\ldots,r$ and so $\{|f_j\>\}_{j=1}^r$ is an orthonormal set. This shows that Eq. (\ref{24}) is valid for
$c_{j}=c''_j$, $|e_j\>=|e''_j\>$ and $|g_j\>=|f''_j\>$. Clearly,  $c_j(j=1,2,\ldots,r)$ are positive numbers that are  independent of $|\psi^{AB}_1\>$, $\{|e_j\>\}_{j=1}^r$ and $\{|g_j\>\}_{j=1}^r$ are orthonormal sets in $\H_A$ and $\H_B$, respectively, that are independent of $|\psi^{AB}_1\>$, and $\{|f_j\>\}_{j=1}^r$ is an orthonormal set in $\H_B$, depending on $|\psi^{AB}_1\>$ and $|\psi^{AB}_2\>$. The proof is completed.

In the same way, one can prove the following result.

{\bf Lemma 2.2.} {\it Let $|\psi^{AB}_1\>$ and $|\psi^{AB}_2\>$ be two purifications of a mixed state $\rho_B$ of system $B$. Then  $|\psi^{AB}_1\>$ and $|\psi^{AB}_2\>$ have the following Schmidt decompositions:
\be\label{L2.2}
|\psi^{AB}_1\>=\sum_{j=1}^r{\sqrt{d_j}}|e_j\>|f_j\>, |\psi^{AB}_2\>=\sum_{j=1}^r{\sqrt{d_j}}|h_j\>|f_j\>
\ee
where $d_j(j=1,2,\ldots,r)$ are positive numbers that are independent of $|\psi^{AB}_1\>$, the sets $\{|h_j\>\}_{j=1}^r$ and $\{|f_j\>\}_{j=1}^r$ are  orthonormal in $\H_A$ and $\H_B$, respectively, and  independent  of $|\psi^{AB}_1\>$, and the set $\{|e_j\>\}_{j=1}^r$ is orthonormal in $\H_A$ and depends  on both $|\psi^{AB}_1\>$ and $|\psi^{AB}_2\>$.}

It was proved in \cite[Lemma 1]{LiBo1} that a set $\{|\Psi_k\>_{AB}\}_{k\in\Gamma}$ of fixed reducing states
can always be written in the following form
$$|\Psi_k\>_{AB}=\sum_{i=1}^d\sqrt{\alpha_i}|i\>_A|b^{(k)}_i\>_B,$$
where $\{|b^{(k)}_i\>_B\}_{i=1}^d$ is a set of orthogonal states. This is indeed a necessary condition for a set $\{|\psi_k\>_A\}_{k\in\Gamma}$ to be masked by an operator $S$.
Generally, it is not sufficient, see Remark 2.4 below.

Due to the impossibility of an arbitrary states
\cite[Theorem 3]{Mask}, it is natural to determine the maskable states of a given physical masker\cite{Li-Wang}. Our following theorem establishes a necessary and sufficient condition for
a set $Q$ in $PS_A$ to be masked by an operator $S$.

{\bf Theorem 2.1.} {\it Let $Q\subset PS_A$ and $S:\H_A\rightarrow \H_A\otimes\H_B$ be a linear operator. Then $Q$ can  be masked by  $S$ if and only if there exist probability distributions (PDs) $\{c_i\}_{i=1}^{r}, \{d_j\}_{j=1}^{r}$ with $c_i>0,d_i>0(i=1,2,\ldots,r)$, and orthonormal sets $\{|e_i\>\}_{i=1}^{r}\subset PS_A$ and $\{|f_j\>\}_{j=1}^{r}\subset PS_B$ such that
\be\label{C2.11}
S|\psi\>=\sum_{i=1}^{r}\sqrt{c_i}|e_i\>|f^\psi_i\>, \<f^\psi_s|f^\psi_t\>=\delta_{s,t}, \ \forall |\psi\>\in Q,\ee
\be\label{C2.12}
S|\psi\>=\sum_{j=1}^{r}\sqrt{d_j}|e^\psi_j\>|f_j\>, \<e^\psi_a|e^\psi_b\>=\delta_{a,b},\ \ \forall |\psi\>\in Q,\ee
where $\delta_{x,y}=0(x\ne y)$ and $\delta_{x,x}=1.$}

 {\bf Proof.} {\it Necessity.} Let $Q$ can  be masked by  $S$. Taking a fixed state $|\psi_0\>\in Q$, we see from  Definition 2.1 that
 \be\label{C2.11-1}
 \tr_B[S|\psi\>\<\psi|S^\dag]=\tr_B[S|\psi_0\>\<\psi_0|S^\dag]:=\rho_A\in D_A,\ \ \forall |\psi\>\in Q,\ee
\be\label{C2.11-2}
\tr_A[S|\psi\>\<\psi|S^\dag]=\tr_A[S|\psi_0\>\<\psi_0|S^\dag]:=\rho_B\in D_B\ \ \forall |\psi\>\in Q.
\ee

For each $|\psi\>\in Q$, using Lemma 2.1 for $|\psi^{AB}_1\>=S|\psi\>$ and $|\psi^{AB}_2\>=S|\psi_0\>$ implies the following Schmidt decompositions with positive coefficients:
 \be\label{C2.11-3}
S|\psi\>=\sum_{j=1}^{r_A}\sqrt{c_j}|e_j\>|f^\psi_j\>,
S|\psi_0\>=\sum_{j=1}^{r_A}\sqrt{c_j}|e_j\>|g_j\>,
\ee
where $\<f^\psi_s|f^\psi_t\>=\delta_{s,t}$ for all $s,t=1,2,\ldots,r_A$, $\{|e_j\>\}_{j=1}^r$ and $\{|g_j\>\}_{j=1}^r$ are orthonormal sets in $\H_A$ and $\H_B$, respectively, independent of $|\psi\>$.

Similarly, for each $|\psi\>\in Q$, using Lemma 2.1 for $|\psi^{AB}_1\>=S|\psi\>$ and $|\psi^{AB}_2\>=S|\psi_0\>$ implies the following Schmidt decompositions with positive coefficients:
 \be\label{C2.11-4}
S|\psi\>=\sum_{j=1}^{r_B}\sqrt{d_j}|e^\psi_j\>|f_j\>,
S|\psi_0\>=\sum_{j=1}^{r_B}\sqrt{d_j}|h_j\>|f_j\>,\ee
where $\<e^\psi_a|e^\psi_b\>=\delta_{a,b}$ for all $a,b=1,2,\ldots,r_B$, $\{|h_j\>\}_{j=1}^r$ and $\{|f_j\>\}_{j=1}^r$ are  orthonormal sets in $\H_A$ and $\H_B$, respectively, independent  of $|\psi\>$.

 Furthermore, the condition (\ref{C2.11-1}) and the first equalities in Eqs. (\ref{C2.11-3}) and  (\ref{C2.11-4}) yield that $$\sum_{i=1}^{r_A}c_i|e_i\>\<e_i|
=\sum_{j=1}^{r_B}d_j|e^\psi_j\>\<e^\psi_j|=\rho_A,$$
and so $r_A=r_B=\dim({\rm{ran}}(\rho_A)),$ denoted it by $r$.
This shows that Eqs. (\ref{C2.11}) and (\ref{C2.12}) hold with the properties that $\<f^\psi_s|f^\psi_t\>=\delta_{s,t}$ and $\<e^\psi_a|e^\psi_b\>=\delta_{a,b}.$

{\it Sufficiency.}  Suppose that Eqs. (\ref{C2.11}) and (\ref{C2.12}) hold with the desired properties. Put
$$\rho_A=\sum_{i=1}^{r}c_i|e_i\>\<e_i|, \rho_B=\sum_{j=1}^{r}d_j|f_j\>\<f_j|,$$
then $\rho_A\in D_A, \rho_B\in D_B$, satisfying
$$
\tr_B[S|\psi\>\<\psi|S^\dag]=\rho_A,
\tr_A[S|\psi\>\<\psi|S^\dag]=\rho_B,\ \forall |\psi\>\in Q.
$$
Thus, $Q$ is masked by  $S$ using Definition 2.1. The proof is completed.

From the proof of the necessity, we find that if $Q\subset PS_A$ and there exists a linear operator $S:\H_A\rightarrow \H_A\otimes\H_B$ such that Eq. (\ref{C2.11-1}) holds, then there exists a  positive probability distribution $\{c_i\}_{i=1}^{r}$ and an orthonormal set $\{|e_i\>\}_{i=1}^{r}\subset PS_A$ satisfying Eq. (\ref{C2.11}) and $\<f^\psi_s|f^\psi_t\>=\delta_{s,t}$. Moreover,
 if $Q\subset PS_A$ and there exists a linear operator $S:\H_A\rightarrow \H_A\otimes\H_B$ such that Eq. (\ref{C2.11-2}) holds,  then there exists a probability distribution  $\{d_i\}_{i=1}^{r}$ and an orthonormal set $\{|f_j\>\}_{j=1}^{r}\subset PS_B$ satisfying
Eq. (\ref{C2.12}) and $\<e^\psi_a|e^\psi_b\>=\delta_{a,b}$.

As an application of Theorem 2.1, we reconsider Example 2.1. From the definition of the operator $S$ in  Example 2.1, we see that for each element
$|\psi\>=\frac{1}{\sqrt{d}}\sum_{k=1}^d{\e}^{{\i}\phi_k}|k\>$ of $Q$, the linearity of $S$ implies that
$$S|\psi\>=\frac{1}{\sqrt{d}}\sum_{k=1}^d{\e}^{{\i}\phi_k}|k\>|k\>
=\sum_{k=1}^d\sqrt{c_k}|e_k\>|f^\psi_k\>,$$
where $c_k=\frac{1}{d}, |e_k\>=|k\>$  and $|f^\psi_k\>={\e}^{{\i}\phi_k}|k\>$, satisfying $\<f^\psi_s|f^\psi_t\>=\delta_{s,t}$, and
$$S|\psi\>=\frac{1}{\sqrt{d}}\sum_{k=1}^d{\e}^{{\i}\phi_k}|k\>|k\>
=\sum_{k=1}^d\sqrt{d_k}|e^\psi_k\>|f_k\>,$$
where $d_k=\frac{1}{d}, |f_k\>=|k\>$ and $|e^\psi_k\>={\e}^{{\i}\phi_k}|k\>$, satisfying $ \<e^\psi_a|e^\psi_b\>=\delta_{a,b}.$ Thus, Theorem 2.1 shows that $Q$ can be masked by $S$ in $\H_A$.

Indeed, the masker used in Example 2.1 is just the operator  $S^\sharp:|k\>\mapsto|kk\>$ considered in \cite{Mask}, which can mask any family of states of the form $\sum_{k=1}^d{\e}^{{\i}\phi_k}
r_k|k\>$ that have the amplitudes $r_k$ in common. Next, we give a generalization of this operator and then derive a characterization of a set that can be masked by the generalized masker $S^\sharp$. To do this, we let $\{|e_k\>\}_{k=1}^d$ and $\{|f_k\>\}_{k=1}^d$ be ONBs for $\H_A$,  and  let $S^\sharp$ be the  isometry from $\H_A$ into $\H_A\otimes\H_A$ with
\be\label{Ssharp}
S^\sharp:|e_k\>\mapsto |e_k\>|f_k\>(\forall k=1,2,\ldots,d).
\ee
With these notations, we have the following result, which gives a generalization of Theorem 4 in \cite{Mask}.

{\bf Corollary 2.1.} {\it A family $Q$ of pure states in $PS_A$ can be masked by $S^\sharp$ if and only if there exists a nonnegative unit vector $\textbf{r}=(r_1,r_2,\ldots,r_{d})^T$ in $\R^{d}$ where $d={d_A}$ such that}
\be\label{C2.1}
Q\subset Q_{\textbf{r}}:=\left\{\sum_{k=1}^d{\e}^{{\i}\phi_k}
r_k|e_k\>:\phi_k\in(-\pi,\pi]\right\}.
\ee

{\bf Proof.} {\it Sufficiency.} Suppose that Eq. (\ref{C2.1}) holds.
For every state $|\psi\>=\sum_{k=1}^d{\e}^{{\i}\phi_k}
r_k|e_k\>$ of  $Q_{{\textbf{r}}}$, we compute that
$$S^\sharp|\psi\>=\sum_{k=1}^d{\e}^{{\i}\phi_k}r_k|e_k\>|f_k\>
=\sum_{k=1}^d\sqrt{c_k}|e_k\>|f^\psi_k\>,$$
where $c_k=r_k^2$  and $|f^\psi_k\>={\e}^{{\i}\phi_k}|f_k\>$, satisfying $\<f^\psi_s|f^\psi_t\>=\delta_{st}$, while
$$S^\sharp|\psi\>=\sum_{k=1}^d{\e}^{{\i}\phi_k}r_k|e_k\>|f_k\>
=\sum_{k=1}^d\sqrt{d_k}|e^\psi_k\>|f_k\>,$$
where $d_k=r_k^2$ and $|e^\psi_k\>={\e}^{{\i}\phi_k}|e_k\>$ with    $ \<e^\psi_a|e^\psi_b\>=\delta_{ab}.$ Thus, Theorem 2.1 shows that $Q_{{\textbf{r}}}$ is masked by $S^\sharp$ and so is $Q$.

{\it Necessity.} Suppose that $Q$ can be masked by $S^\sharp$. Choose a $|\psi_0\>\in Q$ and let $|\psi_0\>=\sum_{k=1}^dr_k{\e}^{{\i}\alpha_k}|e_k\>$ where
$r_k\ge0, \alpha_k\in(-\pi,\pi]$ for all $k$ and $\sum_{k=1}^dr_k^2=1$.
Then we get a nonnegative unit vector $\textbf{r}=(r_1,r_2,\ldots,r_d)^T\in\R^d$.
For every $|\psi\>=\sum_{k=1}^d\lambda_k|e_k\>\in Q$, we have
$S^\sharp|\psi\>=\sum_{k=1}^d\lambda_k|e_k\>|f_k\>,$
and so
$$\tr_B[S^\sharp|\psi\>\<\psi|(S^\sharp)^\dag]=\sum_{k=1}^d|\lambda_k|^2|e_k\>\<e_k|.$$
Especially,
$$\tr_B[S^\sharp|\psi_0\>\<\psi_0|(S^\sharp)^\dag]=
\sum_{k=1}^dr_k^2|e_k\>\<e_k|.$$ Since $Q$ was masked by $S^\sharp$, we obtain   $$\sum_{k=1}^d|\lambda_k|^2|e_k\>\<e_k|=\sum_{k=1}^dr_k^2|e_k\>\<e_k|.$$
This shows that $|\lambda_k|=r_k(k=1,2,\ldots,d)$. Thus, $\lambda_k=
r_k{\e}^{{\i}\phi_k}$ for all $k=1,2,\ldots,d$, where $\phi_k\in(-\pi,\pi]$ is the principle argument of $\lambda_k$ (it is $0$ when $\lambda_k=0$). Hence, $|\psi\>=\sum_{k=1}^dr_k{\e}^{{\i}\phi_k}|e_k\>\in Q_{{\textbf{r}}}$ and so $Q\subset Q_{{\textbf{r}}}$.
 The proof is completed.

{\bf Remark 2.2.} Suppose that $Q$ can be masked by $S^\sharp$. Then we see from Corollary 2.1 that

(1) When $|x\>=\sum_{i=1}^dx_i|e_i\>$ and $|y\>=\sum_{i=1}^dy_i|e_i\>$ are from $Q$, we have $|x_i|=|y_i|(i=1,2,\ldots,d)$.

(2) When $|e_1\>\in Q$, $|e_k\>\notin Q(k\ne 1)$.

(3) For every a nonnegative unit vector $\textbf{r}=(r_1,r_2,\ldots,r_{d})^T$ in $\R^{d}$, the set $Q_{{\textbf{r}}}$ is always masked by  $S^\sharp$.

(4) Fixed a nonnegative unit vector $\textbf{r}=(r_1,r_2,\ldots,r_d)^T\in\R^d$ and an element   $|\psi_0\>=\sum_{k=1}^dr_k{\e}^{{\i}\alpha_k}|e_k\>$ of $Q$,
considering the coordinate  maps
$$P_k\left(\sum_{i=1}^dx_k|e_k\>\right)=x_k(k=1,2,\ldots,d),$$ we observe that for every $|x\>=\sum_{k=1}^dx_k|e_k\>\in Q$, it holds that
$$P_k|x\>\in C_k:=\{z\in\C:|z|=r_k\}(k=1,2,\ldots).$$
Thus, the coordinates $(x_1,x_2,\ldots,x_d)$ of states $|x\>$ in $Q$ under the basis $\{|e_k\>\}_{k=1}^d$ for $\H_A$ belong to the hyperdisk $C_1\times C_2\times\ldots\times C_d\subset\C^d$. If we identify a state
 $|x\>=\sum_{k=1}^dx_k|e_k\>\in Q$ with its coordinates $(x_1,x_2,\ldots,x_d)$, then we can say that maskable states corresponding to the masker $S^\sharp$ belong to the hyperdisk $C_1\times C_2\times\ldots\times C_d$ in $\C^d$. In particular, the maximal set $Q_{{\textbf{r}}}$ of states of $\H_A$ that are masked by $S^\sharp$ is contained in this  hyperdisk.    This  leads to an affirmative answer to Conjecture 5 in \cite{Mask}. Moreover, we observe that when ${{\textbf{r}}_1}$ and ${{\textbf{r}}_2}$ are any two different nonnegative  real unit vectors, both $Q_{{\textbf{r}}_1}$  and $Q_{{\textbf{r}}_2}$  are masked by $S^\sharp$, however their union $Q_{{\textbf{r}}_1}\cup Q_{{\textbf{r}}_2}$ can not be masked by $S^\sharp$.

{\bf Remark 2.3.} When ${{\textbf{r}}}=(r_1,r_2,\ldots,r_d)^T$ is a  real unit vector with $r_i>0(i=1,2,\ldots,d)$, we define
$$\alpha\left(\sum_{k=1}^d{\e}^{{\i}\phi_k}
r_k|e_k\>\right)=(\phi_1,\phi_2,\ldots,\phi_d)^T,$$
and then obtain a bijection $\alpha:Q_{{\textbf{r}}}\rightarrow (-\pi,\pi]^d\subset\R^d$. Thus, states in $Q_{{\textbf{r}}}$ and the points in set $(-\pi,\pi]^d$ are in one to one correspondence. Therefore, we can say that the operator  $S^\sharp$ can mask
 the quantum information encoded in the continuous parameters $(\phi_1,\phi_2,\ldots,\phi_d)\in(-\pi,\pi]^d$. This is just Theorem 4 in  \cite{Mask}.

{\bf Theorem 2.2.} {\it Let $|\psi_1\>,|\psi_2\>\in PS_A$ with $\<\psi_1|\psi_2\>=0$. Then
\be\label{T2.2}
Q=\left\{|\psi_1\>,|\psi_2\>,
\frac{1}{\sqrt{2}}(|\psi_1\>+|\psi_2\>),\frac{1}{\sqrt{2}}(|\psi_1\>-\i|\psi_2\>)\right\}
\ee
can not be masked by a linear operator $S:\H_A\rightarrow \H_A\otimes\H_B$.}

{\bf Proof.} Put
$$|\psi_3\>=\frac{1}{\sqrt{2}}(|\psi_1\>+|\psi_2\>),\ |\psi_4\>=\frac{1}{\sqrt{2}}(|\psi_1\>-\i|\psi_2\>).$$
Suppose that $Q=\left\{|\psi_1\>,|\psi_2\>, |\psi_3\>,|\psi_4\>\right\}$ can be masked by some linear operator $S:\H_A\rightarrow \H_A\otimes\H_B$. Then
Theorem 2.1 implies that there exists a positive PD $\{c_i\}_{i=1}^{r}$ and an orthonormal set $\{|e_i\>\}_{i=1}^{r}\subset PS_A$ such that
\be\label{T2-1}
S|\psi_k\>=\sum_{i=1}^{r}\sqrt{c_i}|e_i\>|f^{\psi_k}_i\>\ (k=1,2,3,4),
\ee
where $\<f^{\psi_k}_s|f^{\psi_k}_t\>=\delta_{s,t}$. Since $Q$ is masked by $S$, we see from Definition 2.1 that there exist states $\rho_A\in D_A$ and $\rho_B\in D_B$ such that
$$
 \tr_B[S|\psi_k\>\<\psi_k|S^\dag]=\rho_A,
\tr_A[S|\psi_k\>\<\psi_k|S^\dag]=\rho_B,\ \forall k=1,2,3,4.$$
Let $(t_1,t_2)=\left(\frac{1}{\sqrt{2}},\frac{1}{\sqrt{2}}\right)$ or
$(t_1,t_2)=\left(\frac{1}{\sqrt{2}},\frac{-\i}{\sqrt{2}}\right)$. Then
$|\psi\>:=t_1|\psi_1\>+t_2|\psi_2\>=|\psi_3\>$ or $|\psi_4\>$, and so
\begin{eqnarray*}
\rho_B=\tr_A[S|\psi\>\<\psi|S^\dag]
=\rho_B+t_1\overline{t_2}\tr_A[S|\psi_1\>\<\psi_2|S^\dag]
+\overline{t_1}{t_2}\tr_A[S|\psi_2\>\<\psi_1|S^\dag].
\end{eqnarray*}
Thus,
$$t_1\overline{t_2}\tr_A[S|\psi_1\>\<\psi_2|S^\dag]
+\overline{t_1}{t_2}\tr_A[S|\psi_2\>\<\psi_1|S^\dag]=0.$$
This shows that
$$\left\{\begin{array}{l}
         \tr_A[S|\psi_1\>\<\psi_2|S^\dag]
+\tr_A[S|\psi_2\>\<\psi_1|S^\dag]=0,\\
         \tr_A[S|\psi_1\>\<\psi_2|S^\dag]
-\tr_A[S|\psi_2\>\<\psi_1|S^\dag]=0.
       \end{array}
\right.
$$
Thus, $\tr_A[S|\psi_1\>\<\psi_2|S^\dag]=0.$
Combining Eq. (\ref{T2-1}), we obtain
\begin{eqnarray*}
0&=&\tr_A[S|\psi_1\>\<\psi_2|S^\dag]\\
&=&\tr_A\left[\sum_{i,j=1}^r\sqrt{c_ic_j}|e_i\>\<e_j|\otimes|f^{\psi_1}_i\>\<f^{\psi_2}_j|\right]\\
&=&\sum_{i=1}^rc_i|f^{\psi_1}_i\>\<f^{\psi_2}_i|.
\end{eqnarray*}
Since $\<f^{\psi_2}_i|f^{\psi_2}_1\>=\delta_{i,1}$, we get
$$0=\left(\sum_{i=1}^rc_i|f^{\psi_1}_i\>\<f^{\psi_2}_i|\right)
|f^{\psi_2}_1\>
=c_1|f^{\psi_1}_1\>,$$
a contradiction. The proof is completed.

Considering the proof of Theorem 2.2 and the observations followed the end of the proof  of Theorem 2.2, we can see that for the set $Q$ in Theorem 2.2, there does not exist a linear operator $S:\H_A\rightarrow \H_A\otimes\H_B$ such that
$$\tr_B[S|\psi_k\>\<\psi_k|S^\dag]=\rho_A,\ \forall k=1,2,3,4.$$

{\bf Remark 2.4} Let $\H_A=\C^2,\H_B=\C^4$, $Q=\{|\psi_k\>\}_{k=1}^4$ where
$$|\psi_1\>=|0\>,|\psi_2\>=|1\>, |\psi_3\>=\frac{1}{\sqrt2}(|0\>+|1\>),
|\psi_4\>=\frac{1}{\sqrt2}(|0\>-\i|1\>).$$ Theorem 2.2 ensures that
$Q$ can not be masked by any $S$.

For $j=1,2$, put
$$|f_j^1\>=|j-1\>|0\>, |f_j^2\>=|j-1\>|1\>,$$
$$
|f_j^3\>=\frac{1}{\sqrt2}(|f_j^1\>+|f_j^2\>)=\frac{1}{\sqrt2}(|j-1\>|0\>+|j-1\>|1\>), $$$$
|f_j^4\>=\frac{1}{\sqrt2}(|f_j^1\>-\i|f_j^2\>)=\frac{1}{\sqrt2}(|j-1\>|0\>-\i|j-1\>|1\>),
$$
then $\{|f_1^k\>, |f_2^k\>\}$ is an orthogonal set of normalized states  for every $k=1,2,3,4$. Define a linear operator $S$ from $\H_A$ into $\H_A\otimes \H_B$ by
$$S|\psi_k\>=\frac{1}{\sqrt{2}}\sum_{j=1}^2|j-1\>|f_j^k\>(k=1,2).$$
Then the linearity of $S$ implies that
$$S|\psi_k\>=\frac{1}{\sqrt{2}}\sum_{j=1}^2|j-1\>|f_j^k\>(k=1,2,3,4)$$
 and so the operator satisfies the condition (2.8) and can not mask $Q$. This shows that Eq. (2.8) (equivalently, Eq. (2) in \cite{LiBo1}) is only a necessary condition for a set $Q$ be to masked.

Since the set $Q$ in Theorem 2.2 can not be masked, we conclude that any set of states in $PS_A$ that contains $Q$ (i.e. $PS_A$) can not be masked. This leads to the following.

{\bf Corollary 2.2}(No-Masking Theorem \cite[Theorem 3]{Mask}). {\it It is impossible to mask the
information in an arbitrary quantum state. That is, $PS_A$ can not be masked.}

Recall that \cite[P. 367]{GuoCao1} a quantum channel $\Phi:B(\H_1)\rightarrow B({\mathcal{H}}_2)$  is said to be {\it trace-type} if it maps all states  as the same state. Equivalently, it is of the form $\Phi(T)=\tr(T)\rho_0$ for a fixed state $\rho_0\in D(\H_2)$  and all $T$ in $B(\H_1)$. An application of the no-masking theorem, we have the following.

{\bf Corollary 2.3.} {\it Let $\dim(\H_A)\ge2$ and   $S:\H_A\rightarrow\H_A\otimes\H_B$ be a linear isometry. Then
one of the quantum channels $\Phi^A_S$ and $\Phi^B_S$ defined by $\Phi^A_S(T):=\tr_B(STS^\dag)$ and   $\Phi^B_S(T):=\tr_A(STS^\dag)$  is not trace-type. That is, there does not exist a pair $(\rho_A,\rho_B)$ of mixed states $\rho_A$ and $\rho_B$  of $A$ and $B$, respectively, such that
\be
\tr_B[STS^\dag]=\tr(T)\rho_A, \tr_A[STS^\dag]=\tr(T)\rho_B,\ \forall T\in B(\H_A).
\ee}

Liang et al. proved in \cite{LiBo2} that the maximal maskable set of qubit states on the Bloch sphere with respect to a fixed qubit state $|p_0\>$ and a masker ${\mathcal{U}}$ is the ones on a spherical circle and then obtained a perfect proof of Conjecture 5 in \cite{Mask}. Similarly, for a state $|\psi_0\>\in PS_A$ and an isometry $S:\H_A\rightarrow\H_A\otimes\H_B$, define
\be
\Omega_S(|\psi_0\>)=\left\{|\psi\>\in PS_A:
\left(
\begin{array}{c}
  \tr_B[|\Psi\>\<\Psi|]\\
  \tr_A[|\Psi\>\<\Psi|]
\end{array}
\right)=\left(
\begin{array}{c}
  \rho_A\\
  \rho_B
\end{array}
\right)
\right\},
\ee
where $|\Psi\>=S|\psi\>$ and
$\rho_A=\tr_B[S|\psi_0\>\<\psi_0|S^\dag], \rho_B=\tr_A[S|\psi_0\>\<\psi_0|S^\dag].$
Clearly, $\Omega_S(|\psi_0\>)$ is masked by $S$ and if $Q\subset PS_A$ is masked by $S$, then $Q\subset \Omega_S(|\psi_0\>)$ for any state $|\psi_0\>$ in $Q$. Hence, $\Omega_S(|\psi_0\>)$ is the largest collection of maskable
states with respect to $|\psi_0\>$ and the linear operator $S$.

\section{Masking of mixed states}
\setcounter{section}{3}
\setcounter{equation}{0}

It was pointed out in \cite[Introduction]{Li-Wang} that if we hide the original quantum information in the mixed states rather than the pure ones, what results can we obtain? We will discuss this question in this section. First, we have to give the definition of a masker for mixed states.

{\bf Definition 3.1.} Let $\mathbb{Q}$ be a subset of $D_A$ and  $S:\H_A\rightarrow \H_A\otimes\H_B$ be a linear operator. If there are mixed states $\rho_A\in D_A,\rho_B\in D_B$ such that
\be\label{3D1}
\tr_B(S\rho S^\dag)=\rho_A{\rm{\ and\ }}
\tr_A(S\rho S^\dag)=\rho_B,\ \  \forall \rho\in \mathbb{Q},\ee
then we say that the information contained in $\mathbb{Q}$ is masked by $S$; shortly, $\mathbb{Q}$ is masked by $S$, or $S$ masks $\mathbb{Q}$.
We also say that  $S$ is a {\it quantum
information masker} for $\mathbb{Q}$ and that $\H_B$ is a {\it masking space} for $\mathbb{Q}$.

When the space $\H_B$ and the operator $S$ satisfy the masking  conditions (\ref{3D1}) exist, we say that $\mathbb{Q}$ can be masked.

See Fig. 2 for the definition above.

 \begin{figure}[H]
\centering
\includegraphics[width=12cm,height=4cm]{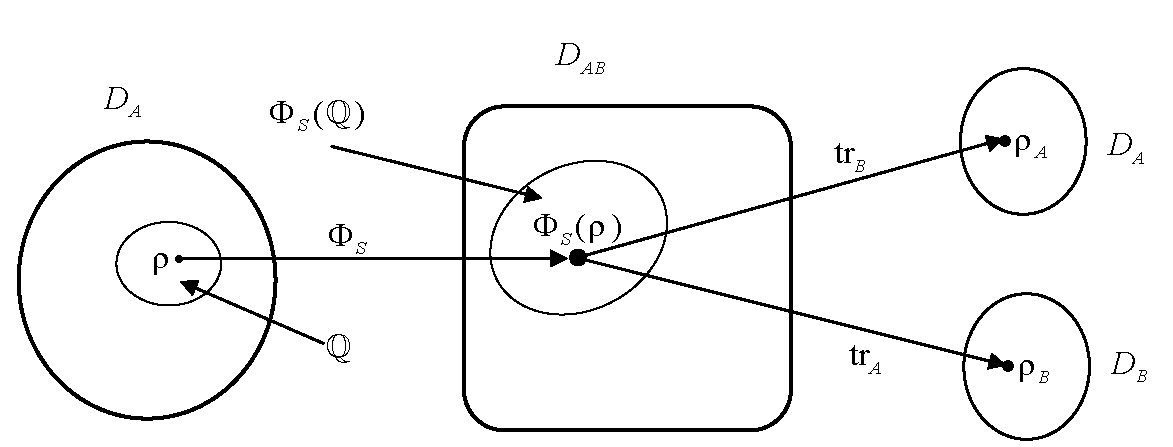}%
\caption{An illustration of Definition 3.1, in which $\Phi_S(T)=STS^\dag$, and all mixed states in $\mathbb{Q}$ are mapped to the fixed states $\rho_A$ and $\rho_B$ under the composite maps $\tr_B\circ\Phi_S$ and $\tr_A\circ\Phi_S$, respectively. Thus, the quantum information contained in the states from $\mathbb{Q}$ is masked by $S$.}
\label{Figure2}
\end{figure}

{\bf Remark 3.1.}
Due to the linearity of $\tr_B\circ\Phi_S$ and $\tr_A\circ\Phi_S$, we see that $\mathbb{Q}$ is masked by $S$ with the marginal states  $\rho_A$ and $\rho_B$ if and only if its convex hull ${\rm{co}}(\mathbb{Q})$ is masked by $S$  with the marginal states  $\rho_A$ and $\rho_B$. Further, when $\mathbb{Q}_k(k=1,2)$ are masked by $S$ with the marginal states  $\rho^{(k)}_A$ and $\rho^{(k)}_B$, the set $p\mathbb{Q}_1+(1-p)\mathbb{Q}_2(0\le p\le1)$ is also masked by $S$ with the marginal states  $p\rho^{(1)}_A+(1-p)\rho^{(2)}_A$ and $p\rho^{(1)}_B+(1-p)\rho^{(2)}_B$. Furthermore, let $Q\subset PS_A$  and let $S:\H_A\rightarrow\H_A\otimes\H_B$ be a linear operator.   Put $\mathbb{Q}=\{|\psi\>\<\psi|:|\psi\>\in\H_A\}$, then $Q$ is masked by $S$ if and only  if $\mathbb{Q}$ is masked by $S$. Thus, Corollary 2.2 yields the following no-masking theorem for mixed states.

{\bf Proposition 3.1.} {\it It is impossible to mask all of mixed states  by a linear operator.}

Thus, a universal masker for all mixed states does not exist and therefore it is useful to find some sets of mixed states that can be masked by an operator. It was proved in \cite{broad[5]} that a commuting set of mixed states can be broadcast. Motivated by this result, we can prove the following conclusion.

{\bf Theorem 3.1.} {\it Any commuting  set $\mathbb{Q}=\{\rho_k\}^{n}_{k=1}$ in $D_A$ can be masked by an isometry $S^{\diamond}:\H_A\rightarrow\H_A\otimes\H_B$ with $\H_B=\H_A$.}

{\bf Proof.} Since $\mathbb{Q}=\{\rho_k\}^n_{k=1}$ is a commuting  set in $D_A$, we see that there is an orthonormal basis $\{|e_1\>,|e_2\>,\ldots,|e_{d_A}\>\}$ for $\H_A$ such that
\be
\rho_k=\sum_{j=1}^{d_A}c^{(k)}_j|e_j\>\<e_j|, \ \forall  k\in[n],
\ee
 where $c^{(k)}_j\ge0 (\forall k\in[n],j\in[d_A])$ satisfying $\sum_{j=1}^{d_A}c^{(k)}_j=1 (\forall k\in[n])$. Let $\omega={\rm{e}}^{\frac{2\pi}{d_A}{\rm{i}}}$ be the basic $d_A$th root of unit. Then $(\omega^p)^{d_A}=1(p\in\Z)$ and so
\be\label{3.3}
1+\omega^p+\omega^{2p}+\cdots+\omega^{(d_A-1)p}=
 \frac{1-(\omega^{p})^{d_A}}
 {1-\omega^{p}}=0
 \ee
for all nonzero integers $p$.

Let $\H_B=\H_A$ and $\{|f_k\>\}_{k=1}^{d_B}$ be any orthonormal basis for $\H_B$, and define
\be\label{3.4}
S^{\diamond}|e_j\>=d_A^{-\frac{1}{2}}
\sum_{k=1}^{d_A}\omega^{(j-1)(k-1)}
 |e_k\>|f_k\>(\forall j\in[d_A]),
\ee
then $S^{\diamond}$ can be extended as a linear operator from $\H_A$ into $\H_A\otimes\H_B$. Eq. (\ref{3.3}) shows that
\bna
&&\tr[S^{\diamond}|e_j\>\<e_m|(S^{\diamond})^\dag]\\
&=&\frac{1}{d_A}
\sum_{k,s=1}^{d_A}\omega^{(j-1)(k-1)-(m-1)(s-1)}
\tr[|e_k\>\<e_s|\otimes|f_k\>\<f_s|]\\
&=&\frac{1}{d_A}
\sum_{k=1}^{d_A}\omega^{(j-1)(k-1)-(m-1)(k-1)}\\
&=&\frac{1}{d_A}
\sum_{k=1}^{d_A}\omega^{(j-m)(k-1)}\\
&=&\delta_{j,m}
\ena
for all $m,j\in[d_A]$.
This shows that $\{S^{\diamond}|e_k\>\}_{k=1}^{d_A}$ is an orthonormal set in $\H_A\otimes\H_B$ and therefore the operator $S^{\diamond}:\H_A\rightarrow\H_A\otimes\H_B$ is an isometry.
Put
$$\rho_A=\frac{1}{d_A}I_A\in D_A, \rho_B=\frac{1}{d_B}I_B\in D_B.$$
For all $k=1,2,\ldots,n$, we see from Eq. (\ref{3.4}) that for each $j=1,2,\ldots,d_A,$
\bna
&&\tr_B[S^{\diamond}|e_j\>\<e_j|(S^{\diamond})^\dag]\\
&=&\frac{1}{d_A}
\sum_{k,s=1}^{d_A}\omega^{(j-1)(k-1)-(j-1)(s-1)}
\tr_B[|e_k\>\<e_s|\otimes|f_k\>\<f_s|]\\
&=&\frac{1}{d_A}
\sum_{k=1}^{d_A}|e_k\>\<e_k|\\
&=&\rho_A.
\ena
Thus,
$$
\tr_B[\Phi_{S^{\diamond}}(\rho_k)]=\sum_{j=1}^{d_A}c^{(k)}_j
\tr_B[S^{\diamond}|e_j\>\<e_j|(S^{\diamond})^\dag]
=\sum_{j=1}^{d_A}c^{(k)}_j\rho_A=\rho_A.$$
Similarly, one can check that $\tr_A[\Phi_{S^{\diamond}}(\rho_k)]=\rho_B$
for all $k=1,2,\ldots,n$. By Definition 3.1, $\mathbb{Q}$ can be masked by $S^{\diamond}$. The proof is completed.

It is remarkable that commutativity of a set $\mathbb{Q}$ of mixed states is only a sufficient condition for masking, but not a necessary one. For example, we consider the set
$$Q_{{\textbf{r}}}:=\left\{\sum_{k=1}^d{\e}^{{\i}\phi_k}
r_k|e_k\>:\phi_k\in(-\pi,\pi]\right\},$$
in Corollary 2.1 where $d\ge2, {{\textbf{r}}}=\left(r_1,r_2,\ldots,r_d\right)$ with $r_k=\frac{1}{\sqrt{d}}(\forall k\in[d])$. For each parameter
$\theta=(\theta_1,\theta_2,\ldots,\theta_d)^T\in (-\pi,\pi]^d,$ define a pure state
$|\psi(\theta)\>=\sum_{k=1}^d{\e}^{{\i}\theta_k}r_k|e_k\>$ and obtain a mixed state $\rho(\theta)=|\psi(\theta)\>\<\psi(\theta)|.$ We see from  Corollary 2.1 that the set $Q_{{\textbf{r}}}$ can be masked by the isometry $S^\sharp$ and then the set  $\mathbb{Q}_{{\textbf{r}}}=\{\rho(\theta): \theta\in (-\pi,\pi]^d\}$ can be masked by ${S^\sharp}$. But $\mathbb{Q}_{{\textbf{r}}}$ is not a commuting set, e.g. when
$$|\psi_1\>=\frac{1}{\sqrt{d}}\sum_{k=1}^d|e_k\>,
|\psi_2\>=\frac{1}{\sqrt{d}}|e_1\>+\frac{1}{\sqrt{d}}
\sum_{k=2}^d{\e}^{\pi{\i}/2}|e_k\>,$$
the corresponding mixed states  $\rho_k=|\psi_k\>\<\psi_k|\in\mathbb{Q}_{{\textbf{r}}}(k=1,2)$ satisfy
$\rho_1\rho_2\ne\rho_2\rho_1.$

Li et al. proved in \cite[Theorem 2]{LiBo1} that every linearly independent set of pure states can be probabilistically masked by a general unitary-reduction operation. As an application of Theorem 3.1, we have the following.

{\bf Corollary 3.1.} {\it Any linearly independent set $Q=\{|\psi_k\>\}_{k=1}^n$ in $PS_A$  can be masked  by an injection  $S:\H_A\rightarrow\H_A\otimes\H_B$.}

{\bf Proof.} Since $Q=\{|\psi_k\>\}_{k=1}^n$ is a linearly independent set, it can be extended as a Hamel basis $Q^*=\{|\psi_k\>\}_{k=1}^{d}$ for $\H_A$ where $d=d_A$. Define a linear operator $T:\H_A\rightarrow \C^d$ by
$$T|x\>=(\<\psi_1|x\>,\<\psi_2|x\>,\ldots,\<\psi_d|x\>)^T, \ \ \forall |x\>\in\H_A,$$ then its adjoint operator reads
$$T^\dag|y\>=\sum_{k=1}^dz_k|\psi_k\>,\ \ \forall |y\>=(z_1,z_2,\ldots,z_d)^T\in\C^d,$$
and so $T^\dag T=\sum_{k=1}^d|\psi_k\>\<\psi_k|.$ Since $Q^*$ is a Hamel basis for $\H_A$, the operator $T^\dag T:\H_A\rightarrow\H_A$ is invertible and so positive-definite. Put $|e_k\>=(T^\dag T)^{-\frac{1}{2}}|\psi_k\>$ for $k=1,2,\ldots,d$, then it is easy to check that  $\{|e_k\>\}_{k=1}^{d}$ becomes an orthonormal basis for $\H_A$. It follows from Theorem 3.1 that the commutative set $\{|e_k\>\<e_k|\}_{k=1}^{d}$ can be masked  by the isometry $S^\diamond:\H_A\rightarrow\H_A\otimes\H_B$ defined by Eq. (3.4).
Thus, $S:=S^\diamond(T^\dag T)^{-\frac{1}{2}}$ is an injection from $\H_A$ into $\H_A\otimes\H_B$ and masks the original linearly independent set $Q$. The proof is completed.

Since a universal masker for all mixed states does not exist (Proposition 3.1), it is important to the sets of mixed states that are masked by a given operator $S$. We call such sets  of mixed states  the maskable sets   of $S$.
Next theorem gives characterization of the maskable sets   of the operator  $S^\sharp$ given by Eq. (\ref{Ssharp}).

{\bf Theorem 3.2.} {\it Let $S^\sharp$ be as in Eq. (\ref{Ssharp}) Then  a set $\mathbb{Q}$ of mixed states in $D_A$ is a maskable set of  ${S^\sharp}$ if and only if there exists a PD vector} $\textbf{p}=(p_1,p_2,\ldots,p_{d})^T$  {\it where $d={d_A}$ such that}
\be\label{3.5}
\Q\subset \Q_{{\textbf{p}}}:=\left\{\rho\in D_{A}: \<e_k|\rho|e_k\>=p_k(\forall k\in[d])\right\}.\ee

{\bf Proof.} {\it Sufficiency.} Suppose that Eq. (\ref{3.5}) holds and define
$$\rho_A=\sum_{k=1}^dp_{k}|e_k\>\<e_k|, \rho_B=\sum_{k=1}^dp_{k}|f_k\>\<f_k|.$$
For every state $\rho=\sum_{k,j=1}^dc_{kj}
|e_k\>\<e_j|$ of  $\Q_{{\textbf{p}}}$, we compute that
$$\Phi_{S^\sharp}(\rho)=\sum_{k,j=1}^dc_{kj}
S^\sharp|e_k\>\<e_j|(S^\sharp)^\dag
=\sum_{k,j=1}^dc_{kj}
|e_k\>|f_k\>\<e_j|\<f_j|,$$
and so
$$\tr_B[\Phi_{S^\sharp}(\rho)]=\sum_{k=1}^dc_{kk}
|e_k\>\<e_k|=\sum_{k=1}^dp_{k}
|e_k\>\<e_k|=\rho_A.$$
Similarly, $$\tr_A[\Phi_{S^\sharp}(\rho)]=\sum_{k=1}^dc_{kk}
|f_k\>\<f_k|=\sum_{k=1}^dp_{k}
|f_k\>\<f_k|=\rho_B.$$
It follows from Definition 3.1 that $\Q_{{\textbf{p}}}$ is masked by ${S^\sharp}$ and so is $\Q$. Thus, $\mathbb{Q}$ is a maskable set of  ${S^\sharp}$.

{\it Necessity.} Suppose that $\mathbb{Q}$ is a maskable set of  ${S^\sharp}$, i.e., $\Q$ can be masked by ${S^\sharp}$. Choose a $\rho_0\in \Q$ and let $p_k=\<e_k|\rho_0|e_k\>$ for $k=1,2,\ldots,d$. Then $\textbf{p}=(p_1,p_2,\ldots,p_d)$ is a PD vector. For every $\rho=\sum_{k,j=1}^dc_{kj}
|e_k\>|e_j\>\in \Q$, we have
$$\tr_B[\Phi_{S^\sharp}(\rho)]=\sum_{k=1}^dc_{kk}
|e_k\>\<e_k|,$$
especially, $\tr_B[\Phi_{S^\sharp}(\rho_0)]=\sum_{k=1}^dp_k=\rho_A.$
Since $\tr_B[\Phi_{S^\sharp}(\rho)]=\tr_B[\Phi_{S^\sharp}(\rho_0)]$, we have  $$\sum_{k=1}^dc_{kk}
|e_k\>\<e_k|=\sum_{k=1}^dp_{k}
|e_k\>\<e_k|,$$
and so $c_{kk}=p_k$ for all $k=1,2,\ldots,d$.
Hence, $\rho\in \Q_{{\textbf{p}}}$ and consequently, $\Q\subset \Q_{{\textbf{p}}}$. The proof is completed.

{\bf Remark 3.2.} Since the set $\Q_{{\textbf{p}}}$ is convex, we see from Theorem 3.2 that $\Q_{{\textbf{p}}}$ is the largest convex set of maskable states $\rho$ satisfying $\<e_k|\rho|e_k\>=p_k(\forall k)$ of ${S^\sharp}$.
Moreover, when two PD vectors $\textbf{p}_1$ and ${{\textbf{p}}}_2$ are not equal, $\Q_{{\textbf{p}}_1}\cup\Q_{{\textbf{p}}_2}$ is not masked by  ${S^\sharp}$, although both $\Q_{{\textbf{p}}_1}$ and $\Q_{{\textbf{p}}_2}$ are  masked by the same ${S^\sharp}$. However, Remark 3.1 yields that a convex combination
$c\Q_{{\textbf{p}}_1}+(1-c)\Q_{{\textbf{p}}_2}$ is masked by   ${S^\sharp}$.

Our last result aims to give the largest maskable set of  ${S^\diamond}$, where $S^\diamond$ is the isometry given by Eq. (\ref{3.4}) in terms of  ONBs $\{|e_k\>\}_{k=1}^{d}$ and $\{|f_k\>\}_{k=1}^{d}$ for $\H_A=\H_B$, that is,
\be\label{3.6}
S^{\diamond}|e_j\>=\sum_{k=1}^{d}
b_{kj}|e_k\>|f_k\>(\forall j\in[d]),
\ee
where $b_{kj}=\frac{1}{\sqrt{d}}\omega^{(j-1)(k-1)}$ with $\omega={\rm{e}}^{\frac{2\pi}{d}{\rm{i}}}.$ Clearly, $|b_{kj}|=\frac{1}{\sqrt{d}}$ for all $k,j\in[d].$

To do this, we write $d=d_A=d_B$ and
\be\label{e}|\varepsilon_k\>=(b_{k1}^{\ *},b_{k2}^{\ *},\ldots,b_{kd}^{\ *})^T,\ee
and use $M_\rho=[\<e_{i}|\rho|e_j\>]$ to denote the matrix of a state $\rho\in D_A$ under the basis $\{|e_k\>\}_{k=1}^{d}$.
Clearly, $\{|\varepsilon_k\>\}_{k=1}^d$ is an ONB for $\C^d$ and so
$$\sum_{k=1}^d\<\varepsilon_k|M_\rho|\varepsilon_k\>=\tr(M_\rho)=1, \ \forall \rho\in D_A.$$

With these notations, we have the following.

{\bf Theorem 3.3.} {\it Let $S^\diamond$ be as in Eq. (\ref{3.6}).  Then  a set $\mathbb{Q}$ of mixed states in $D_A$
 is a maskable set of  ${S^\diamond}$ if and only if there exists a PD vector $\textbf{q}=(q_1,q_2,\ldots,q_{d})^T$ where $d={d_A}$ such that}
\be\label{3.7}
\Q\subset \Q^{{\textbf{q}}}:=\left\{\rho\in D_{A}: \<\varepsilon_k|M_\rho|\varepsilon_k\>=q_k(\forall k\in[d])\right\}.\ee

{\bf Proof.} {\it Sufficiency.} Suppose that Eq. (\ref{3.7}) holds and define
$$\rho_A=\sum_{k=1}^dq_{k}|e_k\>\<e_k|, \rho_B=\sum_{k=1}^dq_{k}|f_k\>\<f_k|.$$
For every state $\rho=\sum_{k,j=1}^dc_{kj}
|e_k\>\<e_j|$ of  $\Q_{{\textbf{p}}}$, Eq. (\ref{3.6}) yields  that
\bna\Phi_{S^\diamond}(\rho)&=&\sum_{i,j=1}^dc_{ij}
S^\diamond|e_i\>\<e_j|(S^\diamond)^\dag\\
&=&\sum_{i,j=1}^dc_{ij}\left(\sum_{k}b_{ki}|e_k\>|f_k\>\right)
\left(\sum_{m}b_{mj}^{\ *}\<e_m|\<f_m|\right)\\
&=&\sum_{i,j}\sum_{k,m}c_{ij}b_{ki}b_{mj}^{\ *}|e_k\>\<e_m|\otimes
|f_k\>\<f_m|.
\ena
From the definitions of $|\varepsilon_k\>$ and $M_{\rho}$, we see that
$$\sum_{i,j}c_{ij}b_{ki}b_{mj}^{\ *}=\<\varepsilon_k|M_\rho|\varepsilon_m\>,\ \ \forall k,m.$$
Thus,
\be\label{3.8}
\Phi_{S^\diamond}(\rho)=\sum_{k,m}\<\varepsilon_k|M_\rho|\varepsilon_m\>\cdot|e_k\>\<e_m|\otimes
|f_k\>\<f_m|.
\ee
Hence,
$$\tr_B[\Phi_{S^\diamond}(\rho)]=\sum_{k=1}^d\<\varepsilon_k|M_\rho|\varepsilon_k\>\cdot
|e_k\>\<e_k|=\sum_{k=1}^dq_{k}
|e_k\>\<e_k|=\rho_A.$$
Similarly, $$\tr_A[\Phi_{S^\diamond}(\rho)]=\sum_{k=1}^d\<\varepsilon_k|M_\rho|\varepsilon_k\>\cdot
|f_k\>\<f_k|=\sum_{k=1}^dq_{k}
|f_k\>\<f_k|=\rho_B.$$
It follows from Definition 3.1 that $\Q^{{\textbf{q}}}$ is masked by ${S^\diamond}$ and so is $\Q$. Thus, $\mathbb{Q}$
 is a maskable set of  ${S^\diamond}$.

{\it Necessity.} Suppose that  $\mathbb{Q}$
 is a maskable set of  ${S^\diamond}$, i.e.,
 $\Q$ can be masked by ${S^\diamond}$. Choose a $\rho_0\in \Q$ and let $q_k=\<\varepsilon_k|M_{\rho_0}|\varepsilon_k\>$ for $k=1,2,\ldots,d$. Then $\textbf{q}=(q_1,q_2,\ldots,q_d)$ is a PD vector. For every $\rho=\sum_{k,j=1}^dc_{kj}
|e_k|e_j\>\in \Q$, Eq. (\ref{3.8}) implies
$$\tr_B[\Phi_{S^\diamond}(\rho)]=\sum_{k=1}^d\<\varepsilon_k|M_\rho|\varepsilon_k\>\cdot
|e_k\>\<e_k|.$$
Particularly, $$\tr_B[\Phi_{S^\diamond}(\rho_0)]=\sum_{k=1}^d\<\varepsilon_k|M_{\rho_0}|\varepsilon_k\>\cdot|e_k\>\<e_k|.$$
Since $\tr_B[\Phi_{S^\diamond}(\rho)]=\tr_B[\Phi_{S^\diamond}(\rho_0)]$, we have  $\<\varepsilon_k|M_{\rho}|\varepsilon_k\>=\<\varepsilon_k|M_{\rho_0}|\varepsilon_k\>=q_k$ for all $k=1,2,\ldots,d$.
This shows that $\rho\in \Q^{{\textbf{q}}}$ and consequently, $\Q\subset \Q^{{\textbf{q}}}$. The proof is completed.

{\bf Remark 3.3.} Since the set $\Q^{{\textbf{q}}}$ is convex, we see from Theorem 3.3 that $\Q^{{\textbf{q}}}$ is the largest convex set of maskable states $\rho$ satisfying $\<\varepsilon_k|M_{\rho}|\varepsilon_k\>=q_k(\forall k)$ of ${S^\diamond}$.
Moreover, when two PD vectors $\textbf{q}_1$ and ${{\textbf{q}}}_2$ are not equal, $\Q^{{\textbf{q}}_1}\cup\Q^{{\textbf{q}}_2}$ is not masked by  ${S^\diamond}$, although both $\Q^{{\textbf{q}}_1}$ and $\Q^{{\textbf{q}}_2}$ are masked by the same  ${S^\diamond}$. However, Remark 3.1 implies that any convex combination  $c\Q^{{\textbf{q}}_1}+(1-c)\Q^{{\textbf{q}}_2}$ is also masked by  ${S^\diamond}$.

When $\mathbb{Q}=\{\rho_k\}^n_{k=1}$ is a commuting  set in $D_A$, there is an ONB $\{|e_k\>\}_{k=1}^{d}$ for $\H_A$ such that
$$
\rho_k=\sum_{j=1}^{d}c^{(k)}_j|e_j\>\<e_j|, \ \forall  k\in[n].
$$
Thus,
$M_{\rho_k}=\sum_{i=1}^dc^{(k)}_j|j\>\<j|$ for all $k\in[n]$, where $\{|j\>\}_{j=1}^d$ is the canonical basis for $\C^{d}$. So,
$$\<\varepsilon_k|M_{\rho_k}|\varepsilon_k\>=\sum_{i=1}^dc^{(k)}_j\<\varepsilon_k|j\>\<j|\varepsilon_k\>
=\sum_{i=1}^dc^{(k)}_j|b_{kj}|^2=\frac{1}{d}$$for all $k\in[n]$. This shows that $\rho_k\in\Q_\textbf{q}$ for all $k\in[n]$ where $q_k=\frac{1}{{d}}(\forall k\in[d])$. Consequently, $\Q\subset\Q_\textbf{q}$. It follows from Theorem 3.3 that $\mathbb{Q}=\{\rho_k\}^n_{k=1}$ can be masked by $\Phi_{S^\diamond}$ in $\H_B=\H_A$, where $S^{\diamond}$ is the isometry given by the basis $\{|e_k\>\}_{k=1}^{d}$ and any orthonormal basis $\{|f_k\>\}_{k=1}^{d}$ for $\H_B=\H_A$ according to Eq. (\ref{3.8}). This gives an alternative proof of Theorem 3.1.

{\bf Remark 3.4.} The quantum Fourier transformation (QFT) ${F}_d=[b_{kj}]=d^{-\frac{1}{2}}[\omega^{(j-1)(k-1)}]$ and the basis $\{|\varepsilon_k\>\}_{k=1}^{d}$ given in Eq. (\ref{e}) has the relationship
$${F}_d^\dag=[|\varepsilon_1\>, |\varepsilon_2\>,\ldots,|\varepsilon_d\>],\
{F}_d=\left[\begin{array}{c}
                \<\varepsilon_1|\\
                \<\varepsilon_2| \\
                \vdots \\
                \<\varepsilon_d|
              \end{array}\right],$$
 and so $[\<\varepsilon_i|M_{\rho}|\varepsilon_j\>]={F}_dM_\rho{F}_d^\dag,$
that is,  $[\<\varepsilon_i|M_{\rho}|\varepsilon_j\>]$ is just the
QFT of the matrix $M_{\rho}$. Thus, the set
$\Q^{{\textbf{q}}}$ in Theorem 3.3 can be rewritten as
\be\label{3.9}
\Q^{{\textbf{q}}}=\left\{\rho\in D_{A}:({F}_dM_\rho{F}_d^\dag)_{kk}=q_k(\forall k\in[d])\right\}.\ee

\section{Summary and conclusions}

In this paper, we have discussed the problem of masking quantum
information encoded in pure and mixed states, respectively, including the mathematical definitions, characterizations, masking theorems, and no-masking theorems. The following conclusions are obtained. (1) An orthogonal (resp. linearly independent) subset of pure states can be masked by an isometry (resp. injection); (2) A commuting subset of mixed states can be masked by an isometry $S^\sharp$; (3)
There are four pure states that are not masked by any operator and then it is impossible to mask all of pure states; (4) It is impossible to mask all of mixed states by an  operator ${S}$. Our discussion leads to an affirmative answer to Conjecture 5 in [Phys. Rev. Lett. 120, 230501 (2018)].

\section*{Acknowledgements}
This subject was supported by the National Natural Science Foundation of China (Nos. 11871318, 11771009) and the Fundamental Research Funds for the Central Universities (GK202007002, GK201903001).


\section*{References}

\begin{thebibliography}{}
\bibitem{Clon[1]} Wootters W K and Zurek W H 1982 A single quantum cannot be cloned \emph{Nature (London)} {\bf299} 802
\bibitem{Clon[2]} Lamas-Linares A, Simon C, Howell J C and Bouwmeester D 2002 Experimental quantum cloning of single photons \emph{Science} {\bf 296} 712¨C714
\bibitem{Clon[3]} Zhang W H, Yu L B, Yang M and Cao Z L 2011 Quantum cloning with multicopy in $d$-dimensions \emph{Sci. China-Phys. Mech. Astro.} {\bf54} 2217
\bibitem{Clon[4]} Guo Z H, Cao H X and Qu S X 2015 Existence and construction of simultaneous cloning machines for mixed states \emph{Sci. China-Phys. Mech. Astro.} {\bf 58} 040302-040302

\bibitem{Clon[5]} Wang M H and Cai Q Y 2019 Duplicating classical bits with universal quantum cloning machine
\emph{Sci. China-Phys. Mech. Astro.} {\bf 62} 030312


\bibitem{broad[5]} Barnum H, Caves C M, Fuchs C A, Jozsa R and Schumacher B 1996 Noncommuting mixed states cannot be broadcast  \emph{Phys. Rev. Lett.} {\bf76}  2818
\bibitem{broad[6]} Kalev A and Hen I 2008 No-broadcasting theorem and its classical counterpart \emph{Phys. Rev. Lett.} {\bf 100} 210502
\bibitem{delt1}  Pati A K and  Braunstein S L 1999 Impossibility of deleting an unknown quantum state \emph{Physics} {\bf 1999}
\bibitem{delt2} Pati A K and Braunstein S L 2000 Quantum no-deleting principle and some of its implications {\emph Physics} {\bf 2000}
\bibitem{hiding[7]} Samal J R, Pati A K and Kumar A 2011 Experimental test of the quantum no-hiding theorem \emph{Phys. Rev. Lett.} {\bf 106}  080401
\bibitem{sign} Barrett J, Hardy L and Kent A 2005 No signalling and quantum key distribution \emph{Phys. Rev. Lett.} {\bf 95}  010503

\bibitem{Mask} Modi K, Pati A K, Sen(De) A and Sen U 2018
Masking quantum information is impossible \emph{Phys. Rew. Lett.} {\bf 120} 230501

\bibitem{[13]} Gisin N, Ribordy G, Tittel W and Zbinden H 2000 Quantum cryptography \emph{Rev. Mod. Phys.} {\bf 74} 145
\bibitem{[14]} Bennett C H, Brassard G, Cr$\acute{\rm{e}}$peau  C, Jozsa R, Peres A and Wootters W K 1993 Teleporting an unknown quantum state via dual classical and Einstein-Podolsky-Rosen channels \emph{Phys. Rev. Lett.} {\bf 70} 1895
\bibitem{[18]} Fiur$\acute{\rm{a}}\breve{\rm{s}}$ek J 2004 Optimal probabilistic cloning and purification of quantum states \emph{Phys. Rev. A} {\bf 70} 032308
\bibitem{18} Gao F, Qin S J, Huang W and Wen Q Y 2019 Quantum private query: A new kind of practical quantum cryptographic protocol \emph{Sci. China-Phys. Mech. Astro.} {\bf 62} 70301

\bibitem{[22]} Zukowski M, Zeilinger A, Horne M and Weinfurter H 1998 Quest for GHZ states \emph{Acta. Phys. Pol. A} {\bf 93}  187

\bibitem{[24]} Cleve R, Gottesman D and Lo H K 1999 How to share a quantum secret \emph{Phys. Rev. Lett.} {\bf83}  648
\bibitem{Li-Wang}Mao-Sheng Li and Yan-Ling Wang, 2018 Masking quantum information in multipartite scenario \emph{Phys. Rev. A} 98, 062306
\bibitem{LiBo1} Li B, Jiang S H, Liang X B, Li-Jost X Q, Fan H and Fei S M 2019 Quantum information masking: deterministic versus probabilistic \emph{Phys. Rev. A} {\bf99} 052343
\bibitem{LiBo2} Liang X B, Li B and Fei S M 2019 Complete characterization of qubit masking \emph{Phys. Rev. A} {\bf 100} 030304(R)

\bibitem{LiMoa}Li Mao-Sheng and Modi Kavan, 2019 Probabilistic and approximate masking of quantum
information, arXiv:1912.02419v1 (2019)
\bibitem{Ding} F. Ding, X. Hu. Masking quantum information on hyperdisks. arXiv:1909.11256 (2019)
\bibitem{LieK}S. H. Lie, H. Kwon, M. S. Kim, and H. Jeong. Unconditionally secure qubit commitment scheme using quantum maskers. arXiv:1903.12304v1 (2019)
\bibitem{LieJ}S. H. Lie and H. Jeong. Randomness cost of masking
quantum information and the information conservation law.
arXiv:1908.07426v1 (2019)
\bibitem{Ghosh}T. Ghosh, S. Sarkar, B. K. Behera, P. K. Panigrahi. Masking of quantum information is possible. arXiv:1910.00938 (2019).
\bibitem{GuoCao1} Guo Z H, Cao H X and Qu S X 2015
Structures of three types of local quantum channels \emph{
Found. Phys.} {\bf 45} 355-369
\end{thebibliography}
\end{document}